\documentclass[a4paper,USenglish,10pt]{article}
\usepackage[utf8]{inputenc}

\usepackage{amsthm,amssymb,amsfonts,amsmath}
\usepackage{amssymb,amsfonts,amsmath}
\newtheorem{theorem}{Theorem}

\theoremstyle{definition}

\usepackage[hidelinks]{hyperref}

\usepackage{graphicx}
\graphicspath{{figures/}}
\usepackage{wrapfig}

\makeatletter
\g@addto@macro\bfseries{\boldmath}
\makeatother

\advance\hoffset-7mm
\usepackage{thm-restate}

\makeatletter
\g@addto@macro\bfseries{\boldmath}
\makeatother

\usepackage{authblk}

\date{}

\usepackage{thm-restate}
\theoremstyle{plain}

\usepackage{xcolor}

\title{Flips in Odd Matchings\footnote{Research on this work has been initiated at the 18th European Geometric Graph Week which was held from September $4^{th}$ to $8^{th}$ 2023 in Alcal\'{a} de Henares. We thank the organizers and all participants for the good atmosphere as well as for inspiring discussions on the topic.}}

\begin{document}
\author[1]{Oswin Aichholzer\footnote{Partially supported by the Austrian Science Fund (FWF) grant W1230.}}
\author[2]{Anna Br\"otzner\footnote{Supported by the Swedish Research Council project ``{\sc illuminate} provably good methods for guarding problems'' (2021-03810).}}
\author[3]{Daniel Perz\footnote{Acknowledges funding by MUR of Italy, under PRIN Project 2022ME9Z78 - NextGRAAL: Next-generation algorithms for constrained GRAph visuALization.}}
\author[4]{Patrick Schnider}

\affil[1]{Graz University of Technology\\
	\texttt{oaich@ist.tugraz.at}}
\affil[2]{Department of Computer Science and Media Technology, Malm\"o University, Sweden\\
	\texttt{anna.brotzner@mau.se}}
\affil[3]{Universit\`{a} degli Studi di Perugia\\
	\texttt{daniel.perz@unipg.it}}
\affil[4]{Department of Computer Science, ETH Z\"{u}rich\\
	\texttt{patrick.schnider@inf.ethz.ch}}
	
	\maketitle

	\begin{abstract}
		Let $\mathcal{P}$ be a set of $n=2m+1$ points in the plane in general position.
		We define the graph $GM_\mathcal{P}$ whose vertex set is the set of all plane matchings on $\mathcal{P}$ with exactly $m$~edges. Two vertices in $GM_\mathcal{P}$ are connected if the two corresponding matchings have $m-1$ edges in common. In this work we show that $GM_\mathcal{P}$ is connected and give an upper bound of $O(n^2)$ on its diameter. Moreover, we present a tight bound of $\Theta(n)$ for the diameter of the flip graph of points in convex position. 
		
	\end{abstract}
	
	\section{Introduction}

	Reconfiguration is the process of changing a structure into another---either through continuous motion or through discrete changes. Concentrating on plane graphs and discrete reconfiguration steps of bounded complexity, like exchanging one edge of the graph for another edge such that the new graph is in the same graph class, a single reconfiguration step is often called an \emph{edge flip}. The \emph{flip graph} is then defined as the graph having a vertex for each configuration and an edge for each flip.
	Flip graphs have several applications, for example morphing~\cite{alamdari2017morph} and enumeration~\cite{avis1996reverse}.
	Three questions are central: studying the connectivity of the flip graph, its diameter, and the complexity of finding the shortest flip sequence between two given configurations.
	The topic of flip graphs has been well studied for different graph classes like triangulations~\cite{amp-fdtsp-15,hurtado1996flipping,kanj2017computing, lawson1972transforming,lubiw2015flip,PILZ2014589,wagner2022connectivity}, plane spanning trees~\cite{bousquet2023reconfiguration,hernando1999geometric}, plane spanning paths~\cite{aichholzer2023flipping,AKL2007}, and many more. For a nice survey see~\cite{survey}.

	For matchings usually other types of flips were considered since a perfect matching cannot be transformed to another perfect matching with a single edge flip.
	A natural flip in perfect matchings is to replace two matching edges with two other edges, such that the new graph is again a perfect matching.
	These flips were studied mostly for convex point sets~\cite{biniaz2019flip,milich2021flips}. While the according flip graph is connected on convex point sets it is open whether this flip graph is connected for any set of points in general position.
	Other types of flips in perfect matchings can be found in~\cite{aichholzer2009compatible,aichholzer2022disjoint,aloupis2013bichromatic}.
	
	In this work we study a setting where single edge flips are possible for matchings.  
	Let~$\mathcal{P}$ be a set of $n=2m+1$ points in the plane in general position (that is, no 3 points are collinear). An \emph{almost perfect matching} on $\mathcal{P}$ is a set $M$ of $m$ line segments whose endpoints are pairwise disjoint and in $\mathcal{P}$\!. The matching $M$ is called \emph{plane} if no two segments cross. 
	
\begin{figure}
		\centering
		\includegraphics[width=0.25\textwidth, page=1]{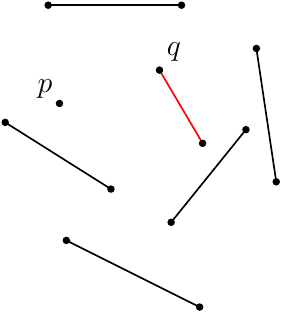}
		\hspace{2.5cm}
		\includegraphics[width=0.25\textwidth, page=2]{figures/flip.pdf}
		\caption{Flipping a matching edge: the previously unmatched point $p$ is matched to~$q$.}
		\label{fig:flip}
	\end{figure}

	Let $\mathcal{M}_\mathcal{P}$ denote the set of all plane almost perfect matchings on $\mathcal{P}$\!. We define the flip graph~$GM_\mathcal{P}$ with vertex set $\mathcal{M}_\mathcal{P}$ through the following flip operation. Consider a matching~$M_1$ and let $p$ be the unmatched point. Let $q \neq p$ be a point in $\mathcal{P}$ such that the segment~$pq$ does not cross any segment in $M_1$. The flip now consists of removing the segment incident to $q$ from the matching and adding $pq$ instead, see \autoref{fig:flip}. Note that this gives another plane almost perfect matching $M_2$. In the graph $GM_\mathcal{P}$, the vertices corresponding to $M_1$ and $M_2$ are adjacent.
	
	In this paper, we prove the following theorem.
	
	\begin{theorem}\label{thm:flip_graph_connected}
		For any set $\mathcal{P}$ of $n=2m+1$ points in general position in the plane the flip graph $GM_\mathcal{P}$ is connected.
	\end{theorem}
	
	In Section~\ref{sec:overview} we give an overview of the used techniques and the proof of \autoref{thm:flip_graph_connected}.
	Then, in Section~\ref{sec:lemma} we prove the lemmata used for the proof of \autoref{thm:flip_graph_connected}.
	In Section~\ref{sec:diameter}, we discuss the diameter of the flip graph, first for points in convex position, and then for general point sets.

	\section{Overview and Proof of \autoref{thm:flip_graph_connected}}\label{sec:overview}
	In this section, we give an overview of our used techniques and the proof of \autoref{thm:flip_graph_connected}. 
	
\begin{figure}
		\centering
		\includegraphics[width=0.98\textwidth]{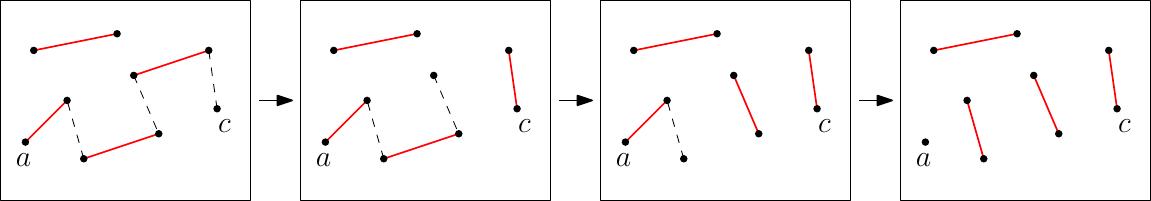}
		\caption{A plane alternating path in the visibility graph gives rise to a sequence of flips.}
		\label{fig:path_to_flip}
	\end{figure}

	Let $G=(V, E)$ be a graph $G$ and let $M$ be a matching in $G$. We call a path $P$ in $G$ an \emph{alternating path} if the edges of $P$ lie alternately in $M$ and in $E \setminus M$. 
	A plane path that alternately consists of matching and non-matching edges and connects to the unmatched point gives rise to a sequence of flips, see \autoref{fig:path_to_flip} for an example. 
	
	To find such a path, we consider so-called \emph{segment endpoint visibility graphs}: graphs that encode the visibility between the endpoints of a set of segments. More precisely, given a set $S$ of (non-intersecting) segments in the plane, its segment endpoint visibility graph is the graph that contains a vertex for every segment endpoint, and an edge between two vertices if the corresponding segment endpoints either (1) are connected by a segment in~$S$, or (2) ``see'' each other, meaning that the open segment between them does not intersect any segment from~$S$. Hoffmann and T{\'o}th~\cite{hoffmann2003segment} proved that segment endpoint visibility graphs always admit a simple Hamiltonian polygon---this is a plane Hamiltonian cycle---, and moreover presented an algorithm to find such a polygon. This result is crucial for us, as a plane perfect matching can be considered as a set of segments in the plane. Hence, for every plane matching $M$ there exists a plane subgraph of the segment endpoint visibility graph of $M$ that is the (not necessarily disjoint) union of a Hamiltonian cycle and $M$\!.
	
	Even disregarding planarity, we show	
	\begin{restatable}{lemmarestatable}{lemmaalternatingpath}
		\label{lem:alt_path}
		Let $G$ be an undirected graph that is the union of a Hamiltonian cycle $C$ and a perfect matching $M$\!.
		Let $e=ab$ be a matching edge and let $c$ be any vertex different from $a$.
		Then there exists an alternating path $P$ that starts with the vertex $a$ and the edge $e$ and ends with the vertex~$c$.
	\end{restatable}

The proof of this lemma, and the one of \autoref{lem:flip_to_hull} stated below, are postponed to Section~\ref{sec:lemma}.

	We denote the \emph{symmetric difference} of two graphs $A,B$ with $A \bigtriangleup B$.
	Given the setup of \autoref{lem:alt_path}, we can compute another matching $M_2 = M \bigtriangleup P$
	in which $a$ is unmatched. 
	This modification corresponds to a sequence of flips in a point set of odd size, see \autoref{fig:path_to_flip}. 
	This flip sequence starts with the matching $M = M_1$ and point $c$ being unmatched, and ends with the matching $M_2$ and point $a$ being unmatched. 
	
	To prove that the flip graph $GM_\mathcal{P}$ is connected, we show that there always exists a sequence of flips that transforms a given plane almost perfect matching into another plane almost perfect matching, where we may choose any fixed point to be the unmatched point. 
	
	\begin{restatable}{lemmarestatable}{lemmaflip}
		\label{lem:flip_to_hull}
		Let $M_1$ be a plane almost perfect matching and let $t$ be an arbitrary point of $\mathcal{P}$\!. Then there exists a sequence of at most $m$ flips to a matching $M_2$ in which the unmatched point is~$t$.
	\end{restatable}

	We use \autoref{lem:flip_to_hull} to show that we can flip every matching $M$ to a \emph{canonical matching}~$M_C$, which we define in the following way. Let $\mathcal{P}=\{p_1,p_2,\ldots,p_{2m+1}\}$, where the points are labeled from left to right. The canonical matching $M_C$ consists of the edges $p_1p_2, p_3p_4,\ldots,p_{2m-1}p_{2m}$ with $p_{2m+1}$ remaining unmatched. It follows from the ordering of the points that this matching is plane.
	
	\begin{proof}[Proof of \autoref{thm:flip_graph_connected}]
		Let $M$ be any plane almost perfect matching on $\mathcal{P}$\!. Let $i \geq 1$ be the smallest index for which the edge $p_{2i-1}p_{2i}$ is not in $M$\!. We show that there is a sequence of flips on the point set $\{p_{2i-1}, p_{2i}, \dots, p_{2m}, p_{{2m+1}}\}$ after which $p_{2i-1}p_{2i}$ is in the resulting matching. In the following, for simplicity of notation, we set $i=1$.
		
		Using \autoref{lem:flip_to_hull}, we first flip to a matching $M_2$ in which the point $p_1$ is unmatched. As the segment $p_1p_2$ is not crossed by any other segment, we perform one more flip which puts $p_1p_2$ into the resulting matching. Now we inductively continue the argument on the point set $\mathcal{P}'=\{p_{3},\ldots,p_{2m+1}\}$ and eventually reach the canonical matching $M_C$. Since thus any matching can be transformed to $M_C$ and because the direction of a sequence of flips can be reverted, the statement of \autoref{thm:flip_graph_connected} follows.
	\end{proof}

	\section{Proofs of the Lemmata}\label{sec:lemma}

	In this section, we prove \autoref{lem:alt_path} and \autoref{lem:flip_to_hull}.
	We begin this section with presenting a procedure to find an alternating path in an abstract graph. Note that we do not require the path to be Hamiltonian. 
	
	\autoref{lem:alt_path} also follows from Lemma~6 in~\cite{hoffmann2003alternatingpath}. For the sake of exposition, we give an alternative proof of the same fact, which is arguably simpler. 
	
\begin{figure}
		\centering
		\includegraphics[width=0.8\textwidth]{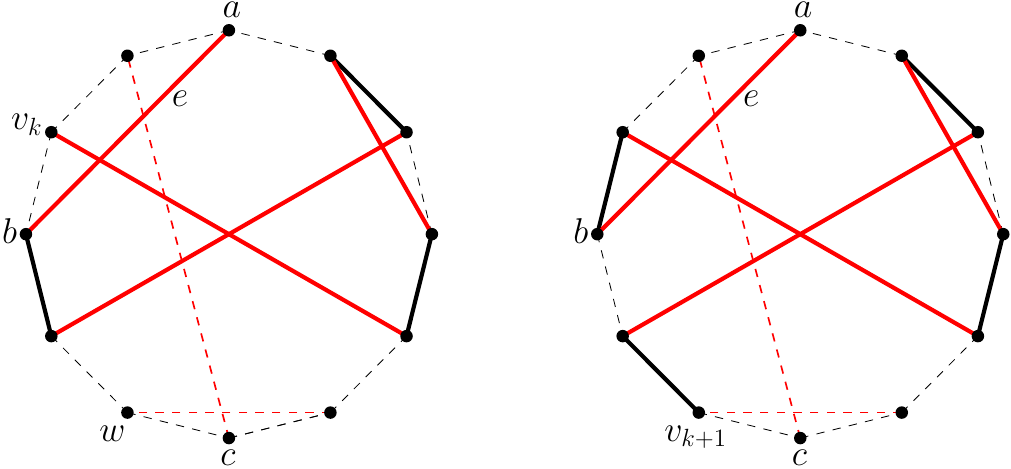}
		\caption{Constructing $G_{k+1}$ (right) from $G_k$ (left). The paths $G_k$ and $G_{k+1}$ are depicted with lines, while unused edges of $G$ are dashed.
			The matching edges are red, the cycle edges are black.}
		\label{fig:path_construction}
	\end{figure}

	\lemmaalternatingpath*
	\begin{proof}
		In a first step, we reduce to the situation where no matching edge except possibly $e$ lies on the cycle $C$, that is, $C\cap M\subseteq\{e\}$. To this end, assume that there is a matching edge $u_1u_2$ lying on the path $\{u_0,u_1,u_2,u_3\}$ of the cycle $C$. We define the graph $G'$ with vertex set $V(G')=V(G)\setminus\{u_1,u_2\}$ by keeping all edges of $G$ induced by $V(G')$ and adding the edge $u_0u_3$.  
		It follows from the construction that $G'$ is again the union of a Hamiltonian cycle and a perfect matching and that $G'$ contains an alternating path starting at $a$ and ending at $c$ if and only if $G$ contains an alternating path starting at $a$ and ending at $c$. As we can iterate this process, in the following we may assume that $C\cap M\subseteq\{e\}$.
		
		We now describe an algorithm that explicitly constructs a required alternating path. The algorithm constructs a sequence of graphs $G_2, G_3,\ldots, G_K$,
		starting with $G_2=\{e\}$, with the following properties:
		\begin{itemize}
			\item[(1)] the graph $G_k$ has $k$ vertices $v_1,\ldots,v_k$;
			\item[(2)] $G_k$ has two vertices of degree 1, namely $v_1$ and $v_k$;
			\item[(3)] all other vertices of $G_k$ have degree 2 and are incident to one edge in $M$ and one edge in $C\setminus M$;
			\item[(4)] $v_1=a$, $v_2=b$ and $v_K=c$.
		\end{itemize}

		From these properties it follows that the last graph $G_K$ is the disjoint union of cycles and the required alternating path $P$\!.
		Let us again point out that we do not require the alternating path to be Hamiltonian. It remains to describe the algorithm and prove that the constructed sequence of graphs satisfies the above properties. We start by setting $G_2=\{e\}$, which trivially satisfies all the properties. In order to construct $G_{k+1}$ from $G_k$ we distinguish two cases, depending on whether in $G_k$ the (unique) edge $\tilde{e}$ incident to $v_k$ is in $M$ or not.
		
		\textbf{Case 1:} $\tilde{e}\in C\setminus M$\!.
		Let $m=(v_k,w)$ be the matching edge incident to $v_k$. We define $G_{k+1}$ by adding $m$ to $G_k$. By Property (3) for $G_k$, all vertices in $G_k$ except $v_k$ are incident to an edge in $M$, and as $M$ is a perfect matching, this implies that $w$ is not a vertex of~$G_k$. Thus, $G_{k+1}$ has one more vertex, proving Property (1) for $G_{k+1}$. The only vertices whose degrees have changed are $w=v_{k+1}$, which now has degree 1, and $v_k$ which is now also incident to an edge in $M$\!. This proves Properties~(2) and~(3).
		
		\textbf{Case 2:} $\tilde{e}\in M$\!.
		For an illustration of this case, see \autoref{fig:path_construction}.
		Consider the unique path~$Q$ in $C$ from $v_k$ to $c$ which does not pass through $a$ and let $w$ be the first vertex on this path that is not a vertex of $G_k$. Let furthermore $Q'$ be the subpath of $Q$ starting at $v_k$ and ending at $w$.
		Set $v_{k+1}=w$. For any edge $e'$ in $Q'$, add $e'$ to $G_{k+1}$ if and only if it is not in $G_k$ and remove it otherwise. Properties (1) and (2) follow directly by definition. For Property~(3), note that the only vertices whose neighborhoods have changed are the vertices on~$Q'$. As $Q'$ is a path on $C$ and we assumed that $C$ contains no matching edge other than $e$, it follows that no matching edge was removed. All vertices are thus still incident to exactly one matching edge. Further, as $C$ is a cycle, every vertex in $Q'$ is incident to exactly two edges in $C\setminus M$. It follows from the construction that exactly one of these edges is removed while the other one is added, proving Property~(3).
		
		Finally, we stop the procedure as soon as we add the vertex $c$, which has to happen for some $G_K$, $K\leq n$, where $n$ is the number of vertices of $G$. This proves the last part of Property~(4) and thus finishes the proof. 	
	\end{proof}
	
	We can now make use of such an alternating path to prove \autoref{lem:flip_to_hull}. 
	
	\lemmaflip* 
	\begin{proof}
		
		Let $p$ be the unmatched point in $M_1$.
		If $p=t$ then we are trivially done, so assume for the remainder that $p \neq t$.
		We duplicate $p$ such that the two points $p, p'$ have the same neighborhood in the segment endpoint visibility graph.
		Moreover, we add the edge $pp'$ to~$M_1$. By Theorem~1 in~\cite{hoffmann2003segment}, there is a plane Hamiltonian cycle $C$ that spans all segment endpoints of $M_1$. Moreover, $M_1 \cup C$ is plane. 
		Let $u$ be the vertex that is matched to $t$ in~$M_1$. 
		By \autoref{lem:alt_path}, there is an alternating path $P$ from $t$ to $p$ in $C \cup M_1$ that starts with the edge~$tu$.
		Since the underlying graph is plane, $P$ is also plane.
		If $p$ and $p'$ are in $P$\!, then the edge $pp'$ is also in $P$ because $pp'$ is a matching edge.
		Hence, we can contract $p$ and $p'$ to a single point $p$ such that $P$ is still an alternating~path.

		Now, we construct a matching $M_2$ by transforming $M_1$ via a sequence of flips along $P$ to get $M_2 = M_1 \bigtriangleup P$\!. 
		$M_2$ is an almost perfect matching in which $p$ is matched, and $t$ is the unmatched point.
		Since $P$ contains at most $m$ matching edges, the sequence consists of at most $m$ flips. 
	\end{proof}
	
	The crux in the proof of \autoref{lem:flip_to_hull} lies in safely duplicating the unmatched point $p$. Both points shall see the same segment endpoints, while visibility between the segment endpoints must not be blocked. This can be achieved by constructing the cell arrangement induced by the lines through all pairs of segment endpoints, and then placing the duplicated point $p'$ in the cell in which $p$ is located. Since the new segment $pp'$ does not intersect any line of the line arrangement, it does not intersect any edge of the segment endpoint visibility graph either. Moreover, any segment endpoint that sees $p$ sees the entire cell in which $p$ is located, and in particular it also sees $p'$\!. 
	
	\section{Diameter of the Flip Graph} \label{sec:diameter}
	Flipping from one plane almost perfect matching $M_1$ to another plane almost perfect matching $M_2$ may take at least linearly many steps. To see this, consider two disjoint matchings $M_1$ and $M_2$. Each flip adds at most one edge of the target matching $M_2$, so it takes at least $m$ flips to transform $M_1$ into $M_2$. 
	
	On the other hand, from the proof of \autoref{thm:flip_graph_connected} it follows directly that no more than $O(n^2)$ flips are needed to transform any plane almost perfect matching on $\mathcal{P}$ into any other plane almost perfect matching on $\mathcal{P}$\!. 
	This is because of the iterative tranformation to the canonical matching $M_C$. By \autoref{lem:flip_to_hull}, it takes up to $m$ flips to transform the matching such that the unmatched point is on the boundary of the convex hull, and one additional flip to add the leftmost edge of $M_C$. Repeating this for every edge of $M_C$ yields the quadratic number of flips. In other words, the diameter of the flip graph $GM_\mathcal{P}$ is in $O(n^2)$.
	
\begin{figure}
		\centering
		\includegraphics[width=0.49\textwidth]{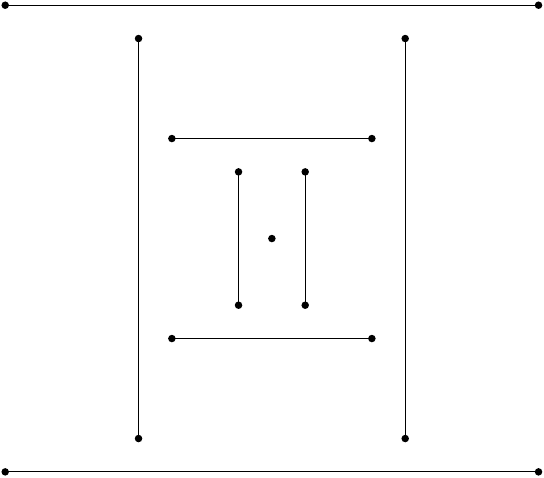}
		\caption{It takes $\Omega(n)$ flips to transform the given matching to any matching where the unmatched point is on the boundary of the convex hull.}
		\label{fig:lower_bound}
	\end{figure}

	\autoref{fig:lower_bound} gives an example where the flip sequence to transfer the unmatched point to the boundary of the convex hull actually has linear length. 
	In each step, the unmatched point gets at most one layer closer to the boundary of the convex hull, thus $\Omega(n)$ steps are necessary. However, we will show that in the specific case of points in convex position, we can do better.

	\begin{theorem}
		For points in convex position, the diameter of the flip graph is $\Theta(n)$.  
	\end{theorem}
	
\begin{figure}
		\centering
		\includegraphics[width=0.9\textwidth, page=1]{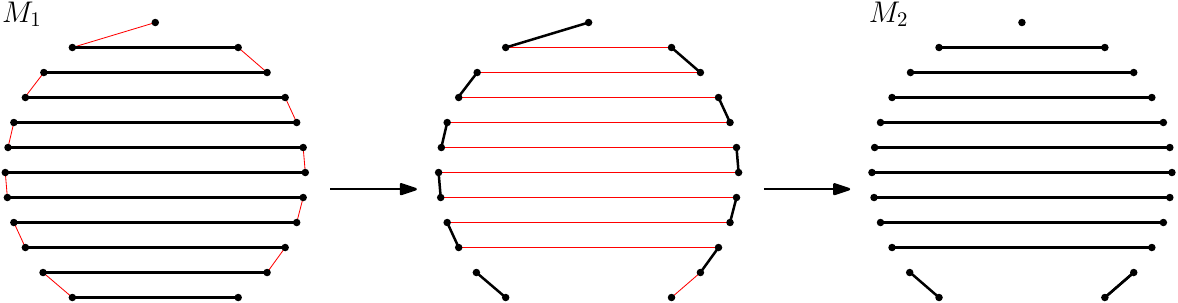}
		\caption{To flip from matching $M_1$ (left, black) to matching $M_2$ (right, black), $m-2$ horizontal edges (all except the two bottom most) need to be flipped twice. The red edges give a shortest flip sequence, which takes $2m$ edge flips.}
		\label{fig:lower_bound_convex_position}
	\end{figure}

	\begin{proof}
	\autoref{fig:lower_bound_convex_position} gives an example for two matchings, which requires $2m$ edge flips to transform one matching into the other, establishing the linear lower bound. 
	
	For the upper bound, we show that any matching can be flipped with $O(n)$ flips to a matching $M_\text{conv}$ with convex hull edges only. Given $M_\text{conv}$, let $e_1, \dots, e_n$ be the matching edges, labeled in counterclockwise direction starting from the unmatched point, and label the end vertices of each edge $e_i$ with $v_i$ and $w_i$, again in counterclockwise direction, and the unmatched vertex with $x$.
	Then, for a given plane almost perfect matching $M_1$, consider the arrangement induced by the boundary of the convex hull and the matching edges. We construct a tree on the dual of this arrangement where every node corresponds to a face, and every edge corresponds to a matching edge. Using this tree structure, we can describe a sequence of edge flips that transforms $M_1$ to a canonical matching $M_\text{conv}$ where each edge is on the boundary of the convex hull. 
	
	We start with the flipping procedure at the face that contains the point $p$ that is unmatched in $M_1$---this face is the root of our tree. Every edge incident to the root encodes a matching edge that can be flipped. 
	There are two different scenarios that specify the type of flip that we choose. Either, $p$ is an endpoint of a matching edge $e_i$ in $M_\text{conv}$ and therefore labeled $v_i$ or $w_i$, or $p$ shall be the unmatched point in $M_\text{conv}$ and is labeled with $x$. If $p$ is labeled $v_i$ or $w_i$, then the edge $e_i$ can always be flipped into the matching since no matching edge can intersect $e_i$ as $v_i$ and $w_i$ are neighboring on the boundary of the convex hull of $\mathcal{P}$\!. 
	After the flip, we update the tree structure and continue with the procedure. The edge $e_i$ connects to a leaf in the new tree and will not be changed anymore.

	If $p$ is labeled $x$, then we consider any child of the root node in the tree that does not connect to a leaf. The edge to this child encodes a matching edge $v_i w_j$ in the current matching, where $i < j$ due to parity reasons. We flip to the edge $x v_i$, leaving $w_j$ as the new unmatched point, update the tree accordingly and continue with the procedure. Since the unmatched point changed, we will be able to add an edge of $M_\text{conv}$ in the next step. 
	
	After at most 2 flips, an edge of $M_\text{conv}$ is added to the matching. It therefore takes at most $2n$ steps to flip from $M_1$ to $M_\text{conv}$. Hence, it takes $O(n)$ flips to transform any plane almost perfect matching on a point set $\mathcal{P}$ in convex position to any other plane almost perfect matching on $\mathcal{P}$\!. 	
	\end{proof}
	
	\section{Conclusion}
	
	We considered the flip graph $GM_\mathcal{P}$ of plane matchings for point sets of odd size, and showed that $GM_\mathcal{P}$ is connected. 
	In the course of the proof, we also showed that the union of a Hamiltonian cycle and a perfect matching always contains an alternating path from an arbitrary matching edge to any arbitrary point.
	
	Moreover, we provided a lower bound of $O(n)$ and an upper bound of $O(n^2)$ on the diameter on the flip graph, as well as a tight bound of $\Theta(n)$ for the situation where the points lie in convex position. It would be interesting to see whether the general upper bound could be improved.

	Finally, the question of connectedness remains open for the flip graph of plane perfect matchings on point sets of even size, where in each flip exactly two edges are replaced. 

\begin{figure}
		\centering
		\includegraphics{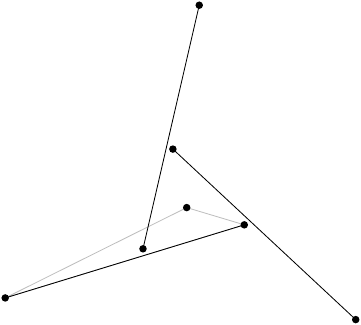}
		\caption{A drawing of an odd matchings such that no empty triangle rotation is possible.}
		\label{fig:no_empty_triangle_rotation}
	\end{figure}

	Another flip operation similar to edge flips is the so-call empty-triangle rotation. That is an edge flip with the additional condition that the added and deleted edge span an empty triangle.
	In other words, an empty-triangle rotation is a more general form of a sweep, where the old edge is rotated until it hits a new point.
	We remark that the flip graph of odd matchings, where a flip is an empty-triangle rotation, is not always connected, as depicted in \autoref{fig:no_empty_triangle_rotation}.

	We also remark that the flip graph (using edge flips) of plane almost perfect bicolored matchings is not always connected where the point set is two colored. For example, this can be achieved by coloring the points in \autoref{fig:no_empty_triangle_rotation} such that the points on the convex hull are red and the other points are blue.

\bibliographystyle{abbrv}

\bibliography{references}

\end{document}